\documentclass[article]{CSL}

\jname{International Journal of Astrobiology}%

\usepackage{amsmath} 
\usepackage{graphicx}
\usepackage{multirow}
\usepackage{booktabs}
\usepackage{natbib}
\usepackage{hyperref} 
\usepackage{graphics} 
\usepackage{subfig} 
\usepackage{xcolor}
\usepackage{enumitem}

\newcommand{\dd}{\mathrm{d}}
\newcommand{\eqn}[1]{Eq.\,(\ref{#1})}

\newcommand{\fig}[1]{Fig.\,\ref{#1}}
\newcommand{\tab}[1]{Tab.\,\ref{#1}}
\newcommand{\case}[1]{\texttt{#1}}
\newcommand{\patmo}{\textsc{Patmo}}
\newcommand{\ith}{$i$th}
\newcommand{\jth}{$j$th}
\newcommand{\repositoryurl}{\url{https://bitbucket.org/patricioavila/patmo_paper/}}

\begin{document}

\supertitle{Research Article}

\title[Chemistry of FFP's exomoons]{Presence of water on exomoons orbiting free-floating planets: a case study}

\author[Patricio Javier \'Avila \textit{et al.}]{Patricio Javier \'Avila$^{1}$, Tommaso Grassi$^{2}$, Stefano Bovino$^{1}$, Andrea Chiavassa$^{3,4}$, Barbara Ercolano$^{2}$, Sebastian Oscar Danielache$^{5,6}$, Eugenio Simoncini$^{6}$}

\address{\add{1}{Departamento de Astronom\'ia, Facultad Ciencias F\'isicas y Matem\'aticas, Universidad de Concepci\'on, Av. Esteban Iturra s/n Barrio Universitario, Casilla 160, Concepci\'on, Chile}; \add{2}{Ludwig Maximilian University of Munich, Scheinerstr. 1, D-81673 Munich, Germany}; \add{3}{Universit\`e C\^ote d'Azur, Observatoire de la C\^ote d'Azur, CNRS, Laboratoire Lagrange, Bd de l'Observatoire, CS 34229, F-06304 Nice Cedex 4, France}; \add{4}{European Southern Observatory, Karl-Schwarzschild-Str. 2, D-85748 Garching bei M\"unchen, Germany}; \add{5}{Department of Material and Life Sciences, Faculty of Science and Technology, Sophia University, 8 Chiyoda, Tokyo, 102-8554, Japan}; \add{6}{Earth and Life Science Institute, Tokyo Institute of Technology, Meguro, Tokyo, 152-8551, Japan};}

\corres{\name{Patricio Javier \'Avila} \email{patricioavila@udec.cl}}

\begin{abstract}
 A free-floating planet is a planetary-mass object that orbits around a non-stellar massive object (e.g.~a brown dwarf) or around the Galactic Center. The presence of exomoons orbiting free-floating planets has been theoretically predicted by several models. Under specific conditions, these moons are able to retain an atmosphere capable of ensuring the long-term thermal stability of liquid water on their surface. We model this environment with a one-dimensional radiative-convective code coupled to a gas-phase chemical network including cosmic rays and ion-neutral reactions. We find that, under specific conditions and assuming stable orbital parameters over time, liquid water can be formed on the surface of the exomoon. The final amount of water for an Earth-mass exomonoon is smaller than the amount of water in Earth oceans, but enough to host the potential development of primordial life. The chemical equilibrium time-scale is controlled by cosmic rays, the main ionization driver in our model of the exomoon atmosphere.
\end{abstract}

\keywords{astrochemistry, atmospheres, planets and satellites.}

\selfcitation{\'Avila PJ, Grassi T, Bovino S,
Chiavassa A, Ercolano B, Danielache SO,
Simoncini E (2021) Presence of water on exomoons orbiting free-floating planets: a case study. \textit{International Journal of Astrobiology} 1--12. https://doi.org/10.1017/
S1473550421000173}

\received{20 February 2021}

\revised{8 April 2021}

\accepted{9 May 2021}

\maketitle

\section{Introduction}
Current planet formation theories predict that during or after the epoch of its formation a planet could be ejected from the hosting planetary system by interacting with a more massive planet \citep{lissauer1987}, or with fly-by stars \citep{laughlin-adams2000}. The concept of starless planets has been introduced several decades ago, hypothesizing that some of them could host life \citep{shapley1958, shapley1962, opik1964}. Although \cite{fogg1990} suggested their existence classifying them as ``singular'' planets or ``unbound'' planets, through decades their definition changed so that most of the studies now employs the term free-floating planet\footnote{Also called rogue, nomad, unbound, orphan, wandering, starless, or sunless planets.} (FFP).

Several attemps of direct observations of possible FFPs are reported (e.g.~\citealt{zapatero-et-al2000, liu-et-al2013, luhman2014, liu2016}) and some of the potential candidates can be observed via microlensing techniques \citep{sumi-et-al2011,bennet-et-al2014,henderson2016,mroz-et-al2018,mroz-et-al2020}. Theoretical estimates show that, on average, in our Galaxy there are two Jupiter-mass \citep{sumi-et-al2011,clanton+gaudi2016} and 2.5~terrestrial-mass free-floating planets per star \citep{barclay-et-al2017}, however, due to the uncertainty on mass, many candidates of FFPs can be classified either as planets or brown dwarves, assuming 13~${\rm M_J}$ the mass threshold between the two classes \citep{caballero2018}.

These free-floating planets are capable of hosting moons. \citet{debes2007} showed that a relevant fraction of terrestrial-sized planets will likely to be ejected while retaining a lunar-sized companion. Exomoons less massive than their host companion are in principle detectable via transit timing variations \citep[see~e.g.][]{teachey-et-al2018}. The ``Hunt for Exomoons with Kepler'' (HEK) project \citep{kipping-et-al2012} aims at searching these natural satellites using data collected by the Kepler space telescope. Despite the uncertainties related to the detection, \cite{teachey+kipping2018} reported the evidence of a potential exomoon, although debated by other authors \citep{rodenbeck2018, heller-et-al2019, kreydberg2019}. The microlensing event studied in \cite{bennet-et-al2014} could be interpreted as a free-floating exoplanet-exomoon system, however, they also point out that this is more probably a\Fpagebreak very-low-mass star with a Neptune-mass planet. Similarly, \cite{fox+wiegert2021} suggested that some systems observed with Kepler are consistent with the presence of dynamically-stable moons, but they also report that no definitive detection can be claimed on this basis, and therefore these systems will require further analysis. The detection limit is determined by the moon-planet mass ratio: for gas giants this is $\sim 10^{-4}$ when formed together (e.g.~in situ formation scenario), that corresponds to an upper limit of a Mars-sized moons \citep{canup+ward2006}, while this could reach Earth-sized objects if captured after a binary exchange encounter \citep{williams2013}, depending on the size and the proximity of the planet and the host star, as well the velocity of the encounter.

The orbital parameters, the characteristics of the hosting star, and the masses of the planet-moon system constrain the habitability of the moon (e.g.~\citealt{heller-et-al2014}). In the case of a moon orbiting a FFP, the absence of stellar illumination suggests that the orbital parameters play the main role, since the orbital eccentricity determines the amount of tidal heating, a key energetic process that may favor the presence of life on the exomoon \citep{reynolds1978, scharf2006, henning-et-al2009, heller2012, heller+barnes2013}.
In addition to that, the thermal budget of the exomoon could be controlled by the evolution of the incident planetary radiation (e.g.~\citealt{haqq-misra2018}), by the runaway greenhouse effect \citep{heller+barnes2014}, and by processes like stellar radiation and thermal heat from the hosting planet \citep{dobos-et-al2017}. Despite these models analyze in detail the role of the various thermal processes, they neglect the effect on the chemistry of the atmosphere, that needs to be considered to determine the conditions for the habitability \citep{lammer-et-al2014}.

Although it has been discussed that a FFP with an atmosphere rich in molecular hydrogen could harbor life \citep{stevenson1999}, it is paramount to model the chemical composition and evolution of CO$_2$ and water to determine the opacity of the atmosphere that might allow liquid water on its surface.
\cite{badescu2010} analyzes four gases (nitrogen, carbon dioxide, methane, and ethane) as the main component of the FFP atmosphere to study the long-term thermal stability of a liquid solvent on the surface, and \cite{badescu2011a,badescu2011b,badescu2011c} studied the possibility for a FFP to host life by calculating the thermal profiles and the solubility properties of condensed gas. It is also shown that these gases produces more effective opacities than H$_2$, as originally suggested by \cite{stevenson1999}.

Overall, these works show that FFP and their moons might represent an environment compatible with the emergency of life within a wide range of masses and different atmospheric compositions, but, to the best of our knowledge, there are no detailed models of the chemical evolution of the atmosphere of a moon orbiting a FFP. Within this context, we introduce here an atmospheric model to tackle this limitation. We assume that in the absence of radiation from a companion star, the tidal and the radiogenic heating mechanisms represent the main sources of energy to maintain and produce an optimal range of surface temperatures. To model the thermal structure of the atmosphere of the exomoon orbiting around free-floating planets, we include these effects in a 1D radiative-convective code\footnote{Code available on \repositoryurl{}} (hereinafter named \patmo{}), alongside a gas-phase chemical kinetics network including cosmic rays, and ion-neutral and neutral-neutral chemistry. We evolve the system in order to determine the chemical evolution in time and the equilibrium time-scale to determine the total amount of water formed.
Since an exhaustive definition of habitability represents a complicated issue \citep{lammer-et-al2009}, we limit ourselves to determine if liquid water is present on the surface of the exomoon (e.g.~\citealt{kasting93}), while changing cosmic ray ionization, chemistry, temperature, and pressure profiles of its atmosphere.

After describing the numerical methods to model the atmosphere we present our results. We then discuss the implications of our findings and future outlooks.

\section{Methods}\label{sec:methods}

\subsection{Planet-moon system}
To model the dynamics of the planet-moon system, we assume that the FFP has been ejected from its host system by e.g.~a perturbation from a gas giant planet, and it retained at least one of its moons after the ejection, as for example discussed in \cite{debes+sigurdsson2007, hong-et-al2018, rabago+steffen2019}. The orbital parameters of the planet-moon system depend on the gravitational interaction during the ejection. \cite{rabago+steffen2019} presented 77~simulations finding that 47\% of the moons are likely to remain bond to the planet, and that the final semi-axis of a survived moon is less than 0.1~au (for comparison Jupiter's largest moons are within 0.01~au), but favoring the innermost orbits, as the number of moons is roughly inversely proportional to the final semi-axis $a$ (namely, the probability of having a closer object after the ejection is higher). The final eccentricity $e$ is mostly between $10^{-3}$ and~1 \citep{hong-et-al2018}. When present, the initial resonance is likely to survive the ejection, allowing the tidal heating to operate on longer time scales compared to the case without resonances, i.e.~10~Myr according to \citealt{heller+barnes2013}. The evolution in time of the orbital parameters determines the temporal domain of our model. In particular, the circularization of the orbit ($e\to0$) plays a key role in reducing the tidal heating produced on the moon by its interaction with the hosting planet, as it will be discussed more in detail in the Section ``Thermal budget'' later in the text. For this reason 10~Myr represents the upper limit of the integration time of our atmospheric model.

We choose a range of semi-major axis and eccentricity following \cite{debes+sigurdsson2007, hong-et-al2018, rabago+steffen2019},  with the former ranging between $0.8\times 10^{-3}$ and $1.4\times 10^{-2}$~au, while the latter between $10^{-3}$ and $5\times 10^{-1}$. We assume a primary object (i.e.~the FFP) with the mass of Jupiter, noting that the stability of the moon-planet system increases with the mass of the hosting planet.

To determine the mass of the moon orbiting around the FFP we follow \cite{barnes+obrien2002} that indicate that a 1~M$_\oplus$ moon is stable when orbiting around a Jupiter-sized planet. \cite{williams2013} show that Earth-sized objects can be captured by a gas giant planet, thus leading to a massive satellite. Since such a planet-moon system could survive an ejection from their stellar system, as reported in the models of \citet{rabago+steffen2019}, we assume a configuration of a secondary object (i.e.~the moon orbiting around the FFP) with a mass of 1~M$_\oplus$ (and gravity $g=980$~cm~s$^{-2}$) orbiting around a 1~M$_{\rm J}$ object, that allows to produce a significant amount of tidal heating to obtain liquid water when the atmosphere is not irradiated by any significant external radiation.

\subsection{Atmospheric modeling}
The chemical composition at different heights of a given exoplanet or exomoon atmosphere is determined by its interplay with the temperature and the vertical pressure profile. In fact, the temperature is driven by the opacity, a function of the chemical composition, while the chemical composition is affected by density and temperature. To model these processes and to predict the relevant quantities we employ \patmo{}, a code aimed at modeling 1D~planetary atmospheres including (photo)chemistry, cosmic rays chemistry, molecular/eddy diffusion, and multi-frequency radiative transfer. \patmo{} employs the \textsc{dlsodes} solver \citep{Hindmarsh2005} to integrate in a full-implicit fashion a set of mass continuity equations
\begin{equation}\label{eq1}
 \frac{\partial n_i}{\partial t} \equiv \partial_t n_i =  P_i-L_i-\frac{\partial \phi_i}{\partial z}\,,
\end{equation}
where $n_i$ is the number density of the \ith{} chemical species, $t$ is time, $P_i$ and $L_i$ are respectively the production and loss rates of the \ith{} species, $\phi_i$ its vertical transport flux, and $z$ is the cell altitude, where $z=0$ represents the surface of the exomoon. We employ 100~linearly-spaced cells ranging from $z=0$ to 35~km and to 30~km, respectively for the low and the high pressure models (see Section ``Results''). The vertical transport flux is
\begin{equation}\label{eq2}
 \phi_i = -K_{zz} n_{\rm tot}\frac{\partial X_{i}}{\partial z}\,,
\end{equation}
where $K_{zz}$ is the eddy diffusion coefficient, $n_{\rm tot}$ the total number density, and $X_i$ the mixing ratio of the \ith{} species, that denotes the relative abundance of the \ith{} species with respect to the total density, i.e.~$X_i=n_{i}/n_{\rm tot}$. The system in \eqn{eq1} allows to solve the time-dependent evolution of each species abundance $n_i$, or when $\partial_t n_i=0$ to find the equilibrium abundances.

As discussed later in this section, $P_i$ and $L_i$ are both functions of the vertical temperature profile, that is directly controlled by the optical depth $\tau$. In principle, to determine the optical depth, we should track the evolution of the radiation when affected by a large number of molecular lines, that depend on the chemical composition and on the temperature, but this will be simplified by using an averaged opacity over the complete spectrum, i.e.~a so-called gray opacity.

When $\tau$ is defined, the temperature of a non-irradiated atmosphere in a radiative thermal equilibrium can be expressed by \citep{marley+robinson2015}
\begin{equation}\label{eq3}
 T(\tau) = T_{\rm eff}\left[\frac{1}{2} (1+D\tau)\right]^{0.25}\,,
\end{equation}
that holds when the atmosphere is in the radiative regime, and where $D=1.5$ is the diffusivity factor, $\tau$ is defined in order to be zero in the outermost layer of our modeled atmosphere and increases toward the surface, and $T_{\rm eff}$ the effective temperature. The diffusivity factor ranges from 1.5 to 2 and depends on the geometrical details of the atmosphere, however, changing its value does not play any crucial role in our findings.

The black-body effective temperature follows the Stefan-Boltzmann law,
\begin{equation}\label{eq4}
 T_{\rm eff}^4 = \frac{\dot{E}_{\rm total}} { 4 \pi \sigma_{\rm sb}\epsilon_{\rm r} R^2 }\,,
\end{equation}
where $\dot{E}_{\rm total}$ is the total flux of energy (see ``Thermal Budget'' subsection), $\sigma_{\rm sb}$ is the Stefan-Boltzmann constant, $\epsilon_{r}=0.9$ is the infrared emissivity factor \citep{henning-et-al2009}, and $R$ is the radius of the moon. The lower zone of the atmosphere is in the convective regime if \citep{sagan1969,weaber+ramanathan1995,robinson+catling2012}
\begin{equation}\label{eq5}
 \frac{\dd\log T}{\dd\log p} > \nabla_{\rm ad}=\frac{\gamma-1}{\gamma}\,,
\end{equation}
where $p$ is the pressure, $\nabla_{\rm ad}$ is the (dry) adiabatic lapse (i.e.~the temperature vertical gradient), and $\gamma$ is the adiabatic index, that depends on the chemical composition. If the atmosphere is in a convective regime, the relation between the pressure $p$ and the temperature $T$ is given by \citep[p. 38]{wallace+hobbs2006}
\begin{equation}\label{eq6}
 T=T_{r}\left(\frac{p}{p_{r}}\right)^{\nabla_{\rm ad}}\,,
\end{equation}
where $T_{\rm r}$ and $p_{\rm r}$ are the temperature and the pressure evaluated at the boundary between the convective and the radiative regimes.
Since by construction at the outermost layer of the atmosphere $\tau=0$, from \eqn{eq3} the temperature depends only on the effective temperature
\begin{equation}\label{eq7}
 T (\tau=0) = 2^{-1/4}T_{\rm eff}.
\end{equation}
We evaluate \eqn{eq5} with finite differences to determine if the atmosphere is in radiative or convective regime, applying \eqn{eq3} or \eqn{eq6} accordingly, i.e.
\begin{equation}\label{eq8}
 T=
    \begin{cases}
     T_{\rm eff}\left[\frac{1}{2} (1+D\tau)\right]^{1/4} & \text{if } \frac{\dd\log T}{\dd\log p} < \nabla_{\rm ad}\\
    T_{r}\left(\frac{p}{p_{r}}\right)^{\nabla_{\rm ad}}  & \text{if } \frac{\dd\log T}{\dd\log p} > \nabla_{\rm ad}\,.
    \end{cases}
\end{equation}

\subsection{Opacity}
The absence of external radiation suggests that a moon orbiting a FFP requires an optically-thick atmosphere to prevent the loss of thermal energy and to keep a temperature that allows liquid water on its surface.
In principle, the opacity of the atmosphere should be calculated dividing the radiation spectrum into several energy bins to cover the key spectral features of the chemical species involved. However, for our aims, this can be reasonably approximated by using an averaged mean opacity over the whole spectrum \citep{hansen2008,guillot2010, robinson+catling2012,parmentier+guillot2014}. We decided to employ the Rosseland mean opacity because of the absence of stellar radiation, i.e.
\begin{equation}\label{eq9}
 \frac{1}{\kappa_{\rm r}}=\int_0^{\infty}\frac{1}{\kappa_v} F(\nu)\, \dd\nu\,,
\end{equation}
where $F(\nu)$ is the weighting function and $\kappa_v$ is the monochromatic opacity of a mixture of gases composed of $N$ chemical species
\begin{equation}\label{eq10}
 \kappa_\nu=\sum_{i=1}^N \kappa_{\nu,i} q_i\,,
\end{equation}
where $\kappa_{\nu,i}$ and $q_i$ are respectively the monochromatic opacity at a given frequency $\nu$, and the mass-specific concentration of the \ith{} species, with $\sum_{i=1}^N q_i = 1$.
The weighting function is
\begin{equation}\label{eq11}
 F(v) = \frac{\frac{\partial B_\nu(T)}{\partial T}}{\int_0^\infty \frac{\partial B_\nu(T)}{\partial T}\dd \nu}\,,
\end{equation}
where $B_\nu(T)$ is the Planck spectral radiance at frequency $\nu$ of a black body with temperature $T$.

From the opacity we obtain the optical depth as a function of pressure,
\begin{equation}\label{eq13}
 \tau (p)=-\int_{z|p}^{z_{\rm top}} \kappa_{\rm r}(z) \rho_a dz=\int_p^{p_{\rm top}} \kappa_{\rm r}(p) g^{-1} dp \,,
\end{equation}
where $\rho_a$ is the mass density of the atmosphere, $g$ is the surface gravity of the moon, $z_{\rm top}$ and $p_{\rm top}$ are respectively the altitude and the pressure of the outermost layer, i.e.~where $\tau=0$, and $z|p$ is the altitude $z$ corresponding to the pressure $p$.
Once the gray opacity is known, we can compute the pressure-dependent temperature vertical profiles. To obtain the Rosseland mean we employ the opacity tables from \cite{badescu2010}, that have been specifically designed for FFPs. These values are valid from 50 to 647.3~K and for pressures between $10^{-5}$ and 2.212$\times10^2$~bar.
Among the possible atmospheric chemical species that compose planetary atmospheres, we exclude N$_2$ because of the small opacity (e.g.~\citealt{badescu2010}), and molecular hydrogen and helium, since we assume that most of their mass has been lost over short time-scales via atmospheric escape, as it often happens in small bodies \citep{catling+zahnle2009}. Other species have higher opacity than molecular hydrogen \citep{stevenson1999, badescu2010}, as for example methane, but planets formation theories indicate that a methane-based atmosphere is unlikely to exist \citep{pollack+yung1980, lammer-et-al2014, massol-et-al2016, lammer-et-al2018}. Ammonia, that has been for example proposed to be the origin of Titan's N$_2$ \citep{Glein2015}, and it has been observed on the Saturnian moons, but not on the Jovian ones \citep{Clark2014, Nimmo2016}, could also play a role as an additional opacity term (e.g.~\citealt{Kasting1982}). Since the amount of ammonia depends on the moon's formation history (e.g.~\citealt{Mandt2014}), in our case we assume that CO$_2$ alone controls the thermal evolution of a FFP atmosphere as it happens in our Solar System, and we therefore employ it in our model, assuming $\gamma=1.3$ \citep{robinson+catling2012}.

\subsection{Thermal budget}\label{sect:thermal_budget}
The total energy flux in \eqn{eq4} is the sum of the tidal and radiogenic heating terms
\begin{equation}\label{eq14}
 \dot{E}_{\rm total} = \dot{E}_{\rm tidal}+\dot{E}_{\rm radio}\,.
\end{equation}
The global heat tidal generation rate $\dot{E}_{\rm tidal}$ of a spin-synchronous homogeneous body on an eccentric orbit, assuming that its stiffness and dissipation are constant and uniform in time, reads  \citep{henning-et-al2009,murray+dermott2000}
\begin{equation}\label{eq15}
 \dot{E}_{\rm tidal}=\frac{21}{2}  \frac{G k_2  M_{\rm p}^2  R_{\rm s}^5  n  e^2  }{Q a^6}\,,
\end{equation}
where $k_2$ is the second-order Love number of the satellite (secondary body), $G$ is the gravitational constant, $M_{\rm p}$ is the mass of the planet (primary body), $R_{\rm s}$ is the radius of the satellite, $n$ is the mean orbital motion, $e$ is the eccentricity, $Q$ is the quality factor of the satellite, and $a$ is the semi-major axis. The masses and radii of the bodies can be determined from planetary formation simulations and from available exoplanetary data. The eccentricity and the semi-major axis are free parameters in our models. In contrast, $k_2$ and $Q$ depend on the material properties, specifically on the complex internal deformation processes. By definition \citep{henning-et-al2009,murray+dermott2000},
\begin{equation}\label{eq16}
 k_2 = \frac{3}{2}\frac{1}{1+\bar{\mu}}\,,
\end{equation}
where $\bar{\mu}$ is the effective rigidity of the body defined as
\begin{equation}\label{eq17}
 \bar{\mu} = \frac{19}{2}\frac{\mu}{\rho_s g R_{\rm s}}\,,
\end{equation}
where $\mu$ is the material rigidity and $\rho_s$ is the mass density of the satellite. We assume common values for rocky bodies, $\mu = 5 \times 10^{11}$~dyne~cm$^{-2}$ and $Q=100$ \citep{yoder+peale1981}.

The tidal heating, as modeled in \eqn{eq15}, depends on the orbital parameters and consequently on their evolution in time. However, in our model we assume that these parameters are constant, but we expect that the circularization of the orbit ($e\to0$) will reduce the total amount of heating in approximately 10~Myr \citep{heller+barnes2013}. For this reason we limit the time-scale of our models to 10~Myr, and we plan to discuss the coupling of the orbital parameters evolution with the atmospheric modeling in a forthcoming work.

The radiogenic heating is determined by the abundances of the radioactive sources similar to those of Earth, hence, we scale it to the total mass of the moon following \citet{henning-et-al2009}. The internal heat flux for the Earth is estimated between $3\times 10^{20}$ to $4.6\times 10^{20}\,\rm{erg \, s}^{-1}$ \citep{jaupart-et-al2007}. About half of this heat, around $2.4\times 10^{20}\,\rm{erg \, s}^{-1}$ \citep{dye2012}, is originated by radioactive decay, while the rest corresponds to other minor sources, including residual formation heat. Although the heat flux could be higher at early times, about 9~times in the early Earth \citep{henning-et-al2009}, we include radiogenic heating for the sake of completeness, but we note that in our models tidal heating is the dominant factor, being at most 2~orders of magnitude larger than the radiogenic heating.

\subsection{Low-temperature chemistry}
The amount of heating produced by the tidal forces, and maintained by the optically-thick atmosphere, produces a relatively low-temperature environment, i.e.~below 300~K (see Section ``Results''), and for this reason we limit our chemical network to include the reactions that are effective in this specific temperature range.

We use a reduced version of the \textsc{STAND2015} chemical network  \citep{rimmer+helling2016}, assuming that photochemistry is not relevant, since by definition a FFP and its satellite do not receive any significant radiation from any companion star. The \textsc{STAND2015} Atmospheric Chemical Network contains H-, C-, N-, and O-based chemical compounds and He, Na, Mg, Si, Cl, Ar, K, Ti, and Fe atoms (necessary in their models for high-temperature chemistry, e.g.~lightning). Since in our case we are interested in a low-temperature environment, we limit our study to reactions including molecules formed by H, C, and O atoms only.  We reduced the total number of species to~101, similarly to \cite{tsai-et-al2017}, and added the reverse rates that are relevant in our temperature range. We also discarded every two-body reaction with a rate constant smaller than $10^{-15}$~cm$^3$~s$^{-1}$ at 300~K and three-body reactions which are only active at temperatures above the ones considered in this work. For the sake of completeness, from \textsc{KIDA} database\footnote{\url{http:\\kida.astrophy.u-bordeaux.fr}} \citep{Wakelam2015} we included cosmic rays reactions not present in \textsc{STAND2015}, but that involve the species in the network. The final network consists of 790~reactions, including three-body, ion-molecule, and cosmic rays chemistry. \patmo{} has been benchmarked with the publicly available code \textsc{vulcan} \citep{tsai-et-al2017}, providing a perfect matching of the chemical vertical profiles. These tests and the chemical network are available in the code repository\footnote{\repositoryurl}.

\subsection{Cosmic-ray chemistry}
Since photochemistry is not included, the main ionization driver are cosmic rays, that consequently determine the indirect formation and the direct destruction of the key molecular species. Cosmic-ray dissociation and ionization are parameterized by using the cosmic-ray ionization rate (CRIR\footnote{Note that CRIR rate $\zeta$ is defined with respect to the ionization of H$_2$ by cosmic rays, but for the sake of simplicity we use the same acronym also when discussing other cosmic rays-driven ionization and dissociation reactions.}) $\zeta$, and we follow \cite{rimmer+helling2013} to compute the atmospheric attenuation of the impinging interstellar cosmic rays from the outermost layer,
\begin{equation}\label{eq18}
 \zeta_j=\zeta_{0}e^{-\sigma N_j}\,,
\end{equation}
where $\zeta_j$ is the CRIR on the \jth{} atmospheric layer, $\zeta_0$ the non-attenuated flux (i.e.~$z\to\infty$), $\sigma$ the collisional averaged cross-section, and $N_j$ the column density obtained integrating the total density from the outermost layer of the atmosphere down to the \jth{} layer. We employ $\sigma=10^{-25}$~cm$^{2}$ \citep{MolinaCuberos2002}, and to mimic the different locations of the hosting planet in the Galaxy, we vary the CRIR between $10^{-17}$ and $10^{-12}$~s$^{-1}$, the latter value to study the effects of a considerably stronger and somewhat extreme case.  We also note that this value could be further reduced from the shielding of the magnetosphere of the hosting planet, as found for example in the Jupiter-Europa system \citep{Nordheim2019}, or as discussed in the time-dependent model by \citet{Heller2013}. These works suggest that in some cases the presence of magnetic fields play a crucial role in determining the amount of cosmic rays (and charged particle in general) impinging the atmpshere of the exomoon, affecting the position of the ``habitable edge'', i.e.~the minimum circumplanetary orbital distance, at which the moon becomes uninhabitable \citep{heller+barnes2013}.

\section{Results}\label{sect:results}
\subsection{Thermal profiles}
\begin{figure}
 \centerline{\includegraphics[width=0.5\textwidth]{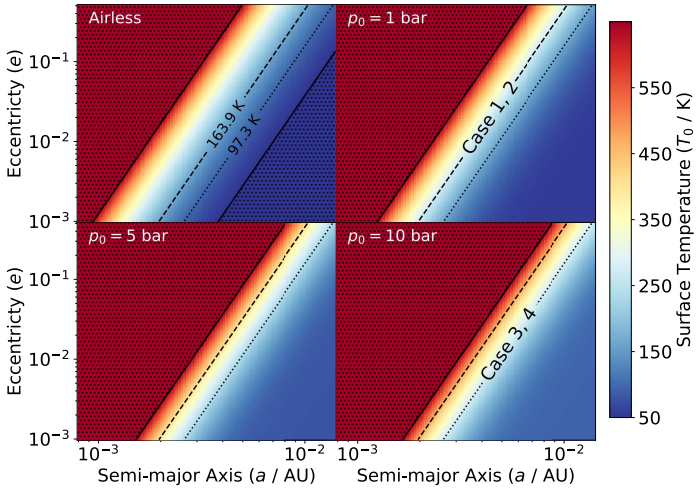}}
 \caption{Surface temperature of a satellite with 1~M$_\oplus$, orbiting around a planet of 1~M$_{\rm J}$. Eccentricity ($e$) ranges between $10^{-3}$ and $5\times 10^{-1}$, while semi-major axis ($a$) between $0.8\times 10^{-3}$ and $1.4\times 10^{-2}$~au. Left top panel: the satellite has no atmosphere (airless body), hence $T_0=T_{\rm eff}$. Other panels: same as first panel, but for different surface pressures $p_0$ assuming a CO$_2$-dominated atmosphere. As reference, the dotted and dashed lines are isothermal contours at respectively $T_{\rm eff}=97.3$ and $163.9$~K assuming the values of the first panel. The latter effective temperature corresponds to a surface temperature of 274.5~K in models \case{Case1} and \case{Case2}, and analogously, the former, in models \case{Case3} and \case{Case4}. The upper and the lower bounds of the colorbar are determined by the limits of the opacity tables, namely $50$ and $647.3$~K. The area outside these limits is indicated by the dotted hatch.}
 \label{fig1}
\end{figure}

We compute the thermal profile for our atmosphere by employing \eqn{eq8} and assuming radiative thermal equilibrium throughout our work. The temperature on the moon surface ($T_0$) depends on the effective temperature $T_{\rm eff}$ (a function of the orbital parameters) and on the surface pressure $p_0$. In \fig{fig1} we show how $T_{\rm eff}$ is affected by the semi-major axis ($a$) and the eccentricity ($e$); the left top panel shows the effective temperature $T_{\rm eff}$, i.e.~the airless moon without the atmosphere, while the other panels show the surface temperature for different surface pressure values, assuming the opacity of a CO$_2$-dominated atmosphere. We note that the isothermal regions follow a power-law relation between $a$ and $e$ (as shown by the black reference lines in \fig{fig1}). In particular, from \eqn{eq15} expanding the mean motion $n$ in terms of $a$ and $e$, we obtain
\begin{equation}
 \dot{E}_{\rm tidal} \propto \left(\frac{G M_p}{a^3}\right)^{1/2}\frac{e^2}{a^6}\,.
\end{equation}

Since increasing $p_0$ globally increases the surface temperature, \emph{for operational purposes} we choose a specific pressure rather than selecting eccentricity and semi-major pairs to obtain different temperatures, and for this reason in our models we change \emph{only} $p_0$ (and consequently $T_{\rm eff}$), instead of varying $a$ \emph{and} $e$. Note that same ($p$, $T_0$) pairs correspond to multiple ($a$, $e$) pairs, suggesting a degeneracy between the orbital parameters and the atmospheric properties (i.e.~different combinations of orbital parameters lead to the same atmospheric properties). This degeneracy is apparent, being determined by the time-independent modeling of the orbital parameters $a$ and $e$, and therefore it represents only an operational method to reduce the number of atmospheric models and to have a clearer discussion of our results. However, in future studies, where the chemical-atmospheric model will be evolved alongside the orbital parameters, this apparent degeneracy will be removed by construction.

Our parameter space is represented by four limiting cases with two different surface pressure values (1 and 10~bar) and two different cosmic rays fluxes, namely, low ($1.3\times 10^{-17}$~s$^{-1}$) and high ($10^{-12}$~s$^{-1}$), as reported in \tab{tab1}.  Our choice of parameters guarantees a surface temperature for which water is liquid, i.e.~in each case we always obtain $T_0\sim 274.5$~K,  that corresponds to 1.35~times the freezing point in our pressure range.
The two different atmospheric configurations of our 1~M$_\oplus$ exomoon are determined by its formation history and the subsequent evolution \citep{lammer-et-al2018}, and for this reason the mass of the moon has no direct influence on the mass of its atmosphere and on the pressure at the surface, as shown for example by comparing the Earth ($p=1$~$p_\oplus$, $m=1$~M$_\oplus$), Venus (91~$p_\oplus$, 0.82~M$_\oplus$), and Titan (1.5~$p_\oplus$, 0.02~M$_\oplus$). We therefore assume that the two values employed for the surface pressure are plausible and that are unaltered during the simulation.
The lower limit of the cosmic rays flux is typically used in the dense interstellar medium \citep{dalgarno2006}. The upper limit represents an extreme case found in the interstellar gas nearby a gamma-ray-emitting supernova remnants \citep{becker2011}. Although this extreme value reproduces a very specific environment, we choose this as an upper limit that allows to compare the effects of the atmospheric attenuation on the chemistry in the high pressure cases.

\begin{table}
    \tabcolsep4pt
    \processtable{Parameter space explored in this work. The surface temperature $T_0$ in every case is 274.5~K. The last three columns indicate the case description employed in the text for pressure, cosmic rays ionization, and temperature, e.g.~\case{Case1} is low-pressure, low-CRIR, and high-temperature. \label{tab1}}
    {\begin{tabular}{@{\hspace*{4pt}}crrrccc@{\hspace*{4pt}}}
    \rowcolor{Theadcolor}
        Case & $p_0\, [\rm{bar}]$ & $\zeta\, [\rm{s}^{-1}]$ & $T_{\rm eff}\, [\rm{K}]$ & pressure & CRIR & temp \\\hline
        1 & 1$\,\,$ & $1.3\times 10^{-17}$ & 163.9 & low & low & high\\\hline
        2 & 1$\,\,$ & $10^{-12}$ & 163.9 & low & high & high\\\hline
        3 & 10$\,\,$ & $1.3\times 10^{-17}$ & 97.3 & high & low & low\\\hline
        4 & 10$\,\,$ & $10^{-12}$ & 97.3 & high & high & low\\\hline
    \end{tabular}}{\begin{tablenotes}
    \end{tablenotes}}
\end{table}

\begin{figure}
 \centerline{\includegraphics[width=0.5\textwidth]{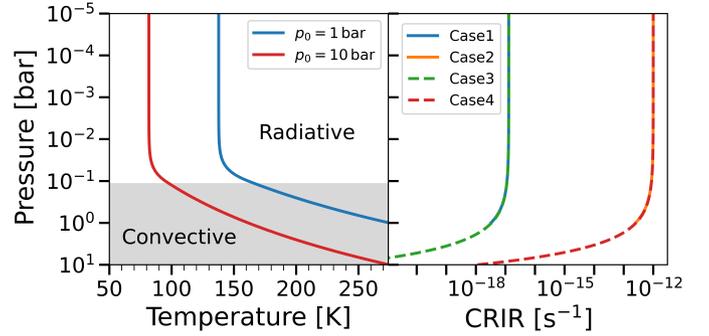}}
 \caption{Left panel: vertical thermal profiles for $p_0=1$~bar (\case{Case1} and \case{Case2}) and 10~bar (\case{Case3} and \case{Case4}). The surface temperature $T_0=274.5$~K is the same in both cases. Thermal profiles are assumed to be constant with time and independent from the CRIR. The gray-shaded area indicates the convective regime. Right panel: cosmic-rays ionization rate (CRIR) as a function of the vertical pressure. Note that \case{Case1} (solid blue) and \case{Case3} (dashed green)  as well as \case{Case2} (solid orange) and \case{Case4} (dashed red) overlap, since the non-attenuated ionization rate is the same. The higher pressure reached by \case{Case3} and \case{Case4} allows the cosmic rays to penetrate deeper in the atmosphere, but with a larger attenuation when reaching the surface.}
 \label{fig2}
\end{figure}

Since we ignore the heating from cosmic rays, being at most an order of magnitude smaller than the radiogenic heating, the vertical thermal profiles reported in \fig{fig2} depend only on $p_0$. Left panel of \fig{fig2} shows the variation of the temperature with pressure (and hence with height) for two models with two different ground pressures ($p_0$), but with the same ground temperature ($T_0$). For this reason in both cases the two thermal profiles guarantee liquid water on the surface.
The vertical thermal profiles are obtained from \eqn{eq3} and \eqn{eq7} by constraining the effective temperature $T_{\rm eff}$, with the optical depth in \eqn{eq3} computed from \eqn{eq13} by constraining the surface pressure $p_0$. In our model the thermal profiles do not change in time, since we assume that the heating sources are constant, and that the atmosphere is always in hydrostatic equilibrium, i.e.~readjustments of the vertical density profile are instantaneous.

The optical depth found in our profiles increases from the outermost layers of the atmosphere ($\tau=0$) reaching $\tau=50.33$ at 1~bar, and $\tau=4767.57$ at 10~bar. For $p\gtrsim10^{-4}$~bar this can be generalized with the power-law $\tau(p) = 50.33\,p^{1.97}$, where the pressure $p$ is in units of bar.

In the right panel of \fig{fig2} we report the vertical profile of the CRIR, for the different models. We note that the attenuation depends on the unaffected ionization, i.e.~$\zeta_0$, and the depth of the atmosphere (i.e.~the pressure reached at the surface). In fact, in the high-pressure models (\case{Case3} and \case{Case4}, surface at $p=10$~bar) the cosmic rays penetrate deeper in the atmosphere, but their attenuation on the surface is larger when compared to the low-pressure models (\case{Case1} and \case{Case2}, surface at $p=1$~bar). In terms of CRIR, the right panel of \fig{fig2} also indicates that \case{Case3} (\case{Case4}) represents an extension of \case{Case1} (\case{Case2}), where the cosmic rays reach higher-pressure layers.

All the models assume that the atmosphere of the moon is initially vertically uniform and composed of 90\%~CO$_2$ and 10\%~H$_2$ (in number density), following a scenario where carbon dioxide is generated by evaporation from rocks, and locked there during the solidification phase when the body was still forming \citep{pollack+yung1980, lebrun-et-al2013, lammer-et-al2014, massol-et-al2016, lammer-et-al2018}.
Under our assumptions molecular hydrogen is a key ingredient for the formation of water, being the initial reservoir of hydrogen, and unlike warmer planets that are known to lose H$_2$ over relatively short periods because of the thermal escape \citep{catling+zahnle2009}, a moon with a cold environment could retain a fraction of molecular hydrogen consistent with the abundances employed in this work \citep{stevenson1999}. This initial abundance of molecular hydrogen is not relevant to affect the opacity, but crucial for the formation of water.

\subsection{Time evolution of water formation}
\begin{figure}
\centering
\includegraphics[width=0.5\textwidth]{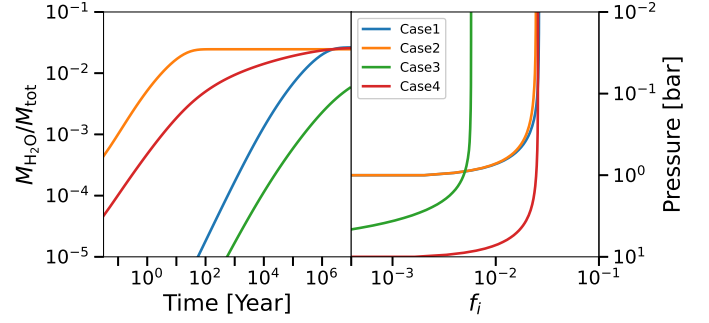}
\caption{Left panel: evolution of the vertically-integrated fraction of water relative to the total mass of the atmosphere, for the models reported in \tab{tab1} as a function of time. \case{Case1} (blue) corresponds to high-temperature, low-pressure, low-CRIR; \case{Case2} (orange)  high-temperature, low-pressure, high-CRIR; \case{Case3} (green) low-temperature, high-pressure, low-CRIR; \case{Case4} (red) low-temperature, high-pressure, high-CRIR. The atmosphere is initially composed of 90\% CO$_2$ and 10\% H$_2$. Right panel: cumulative mass of water  at 10~Myr integrated from the surface as a function of the pressure, i.e.~the \ith{} layer has ${f_i=\sum_{j=1}^{i}M_{{\rm H_2O}, j} / M_{\rm tot}}$. Note that for \case{Case 1} and \case{Case 2} the surface pressure is $p_0=1$~bar, while $p_0=10$~bar for \case{Case 3} and \case{Case 4}.}
\label{fig3}
\end{figure}

We evolve the chemistry in time for 10~Myr according to \eqn{eq1} and we report in the left panel of \fig{fig3} the total amount of water integrated over all layers.
In particular, from the water mixing ratio of each layer $x_{{\rm H_2O},i}$ we compute the total mass of water produced in the atmosphere as ${M_{\rm H_2O}=\mu_{\rm H_2O}\sum_i\Delta V_i n_{{\rm tot},i} x_{{\rm H_2O},i}}$, where the sum is over every layer, $\Delta V_i=4\pi\left(\hat z_{i+1}^3 - \hat z_i^3\right)/3$ is the volume of the \ith{} layer, $n_{{\rm tot},i}$ its total number density, $\mu_{\rm H_2O}$ the mass of a water molecule, and $\hat z_i = z_i + R_{\rm s}$, i.e.~the altitude including the radius of the exomoon. Analogously, the total mass of the atmosphere is ${M_{\rm tot}=\sum_i\mu_i\Delta V_i n_{{\rm tot},i}}$, where $\mu_i$ is the mean molecular weight computed in the \ith{} layer.

The total amount of water formed in \case{Case1} and \case{Case2} is respectively ${M_{\rm H_2O}=1.32\times10^{17}}$~kg (at $t\gtrsim100$~Myr) and $1.24\times10^{17}$~kg ($t\gtrsim100$~yr), while for \case{Case3} and \case{Case4} is $2.85\times10^{17}$~kg and $1.27\times10^{18}$~kg (both at $t\simeq 10$~Myr), respectively.
For comparison, the total mass of water in the Earth's oceans is $\sim 1.4 \times 10^{21}$~kg \citep{weast1979}.
The total corresponding atmospheric mass is ${M_{\rm tot}=5.02\times10^{18}}$~kg (\case{Case1} and \case{Case2}) and ${M_{\rm tot}=4.93\times10^{19}}$~kg (\case{Case3} and \case{Case4}). Again, for comparison, the mass of the Earth's atmosphere is $5.15\times10^{18}$~kg \citep{Trenberth2005}.

The cumulative contribution to the total mass of water at the end of the simulation integrated from the surface layer as a function of the pressure is reported in the right panel of \fig{fig3}. For \case{Case1} and \case{Case2} the surface layers ($p\gtrsim 0.1$~bar) comprise almost the total mass of water present in the atmosphere. Analogously, \case{Case3} and \case{Case4} reach almost the total mass of water at approximately 1~bar, even if \case{Case3} is far from the chemical equilibrium (see left panel of \fig{fig3}).

In \fig{fig3} all the models have a steady increase of the fractional water content over time (note the logarithmic scale), followed by a relative reduction of the formation rate (and in some cases an equilibrium plateau) that depends on the amount of CRIR and on the temperature profile of the specific model. The final fractional amount of water is similar in each model, but reached with different time-scales. A shorter time-scale is found in the models with higher CRIR, i.e.~\case{Case2} and \case{Case4}, being the cosmic rays attenuation by the atmosphere ineffective and thus leading to a faster chemistry, mainly due to the dissociation of the initial reservoir of CO$_2$ and H$_2$. In these models, the chemical kinetic driven by the cosmic rays is effective in every layer, including the deeper ones. The less prominent plateau of the high-pressure \case{Case4} is caused by the missing contribution of the deeper layers, where the CRIR is comparably less effective (see \fig{fig2}).  Conversely, in \case{Case1}, where the CRIR efficiency is smaller, the equilibrium time-scale for water is considerably longer, being around $10^7$~yr, and in the high-density \case{Case3} the equilibrium is not reached even within the simulation time. This slower (and not CRIR-driven) evolution is mainly controlled by the low-temperature chemistry, which is less effective when compared for example to Earth.

The stability observed in \case{Case2} (but also in \case{Case1} and \case{Case4}) is mainly due to the balance between the formation channel ${\rm H_2 + OH \to H_2O + H}$ and the destruction reaction ${\rm H_2O + CR \to OH + H}$ (where CR indicates that the reactant interacts with cosmic rays), as discussed in detail in Section ``Chemical vertical profiles''.
This scenario might be altered in the surface layer by the destruction of water driven by condensation/rain-out processes, i.e.~the removal from the gas phase \citep{hu-et-al2012}, but this process is not included in the present model.

The opacity of the atmosphere is controlled by CO$_2$, the initial oxygen reservoir. This is mainly eroded by the CRIR-driven dissociation reaction ${\rm CO_2 + CR \to CO + O}$, efficiently balanced by the formation routes ${\rm CO + OH \to CO_2 + H}$ and (less relevant) ${\rm O + HCO \to CO_2 + H}$. Since these formation processes are effective during the evolution of every model, CO$_2$ remains relatively constant in time when compared to water, that never becomes dominant over carbon dioxide, causing our approximation of a CO$_2$-dominated atmosphere to hold. Nevertheless, water contributes to the total opacity, as reported by \cite{lehmer-et-al2017}, that explored water-dominated atmospheres of moons around gas giant planets. Even if CO$_2$ is partially replaced with water, due to the high optical depth, the lower part of the atmosphere remains in a convective regime (see \fig{fig2}), and the surface remains warm enough to have liquid water. For comparison, on the Earth (around $10^{-3}$~bar and 220~K) the rotation bands of water vapor play a role for wavelengths $\gtrsim20$~$\mu$m and $\lesssim7.5$~$\mu$m, while CO$_2$ is known to reduce the escaping radiation flux around 15~$\mu$m, where H$_2$O is less efficient \citep{Zhong2013}.

\subsection{Chemical vertical profiles}
\begin{figure}
 \centerline{\includegraphics[width=0.5\textwidth]{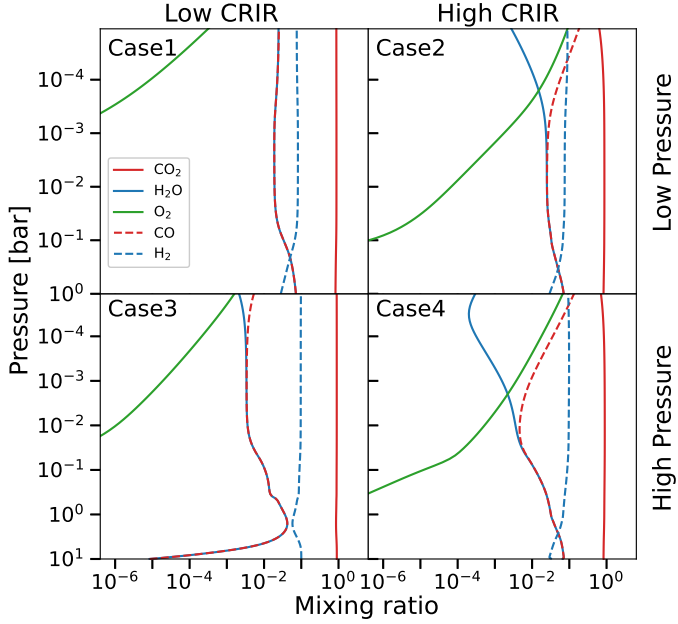}}
 \caption{Vertical distribution of different chemical species for an atmosphere composed of 90\% CO$_2$ and 10\% H$_2$  at $10^7$~yr. Upper panels correspond to high temperature and low pressure, lower panels to low temperature and high pressure cases. Leftmost panels are lower cosmic rays ionization, while rightmost panels higher cosmic rays ionization models. Lines indicate CO$_2$ (solid red), water (solid blue), molecular oxygen (solid green), CO (dashed red), and H$_2$ (dashed blue).}
 \label{fig4}
\end{figure}

Our code not only allows to compute the evolution of the total abundances, but also their vertical profiles as reported in \fig{fig4} for the four cases at $10^7$~yr and for different chemical species (note that e.g.~\case{Case2} reaches this profile after $10^2$~yr already, see \fig{fig3}). As previously discussed, water abundance is determined by the main formation channel ${\rm H_2 + OH \to H_2O + H}$ that reaches equilibrium with ${\rm H_2O + CR \to OH + H}$. The oxygen necessary for the formation of OH, is produced by the reaction ${\rm CO_2 + CR \to CO + O}$, i.e.~controlled by the efficiency of cosmic rays to penetrate the atmosphere of the exomoon (cfr.~\fig{fig2}). In fact, \case{Case3} is the only model presenting a significant variation of water abundance in the surface layers. This specific model has low-CRIR (as \case{Case1}) and a denser atmosphere (as \case{Case4}), hence the chemistry is affected by the noticeably effective CRIR attenuation.

The reduced amount of water in the upper part of the atmosphere in the high-CRIR models (\case{Case2} and \case{Case4}) is determined by the more efficient destruction by CRIR. The difference between these two cases is enhanced by the corresponding temperature profiles, since the reaction ${\rm H_2 + OH \to H_2O + H}$ is more efficient at higher temperatures, i.e.~\case{Case2} produces more water being comparably warmer, see \fig{fig2}.

In the lower layers of every model the presence of the three-body reaction ${\rm H + CO + M \to HCO + M}$ (where M~is a catalyzing species) is the starting point of an additional route for the formation of water. In fact, it favors the formation of CH$_2$O, via ${\rm HCO + HCO \to CH_2O + CO}$, that reacts with H$_3$O$^+$ to form water by ${\rm CH_2O + H_3O^+ \to CH_3O^+ + H_2O}$, this one balanced by its reverse reaction. This process is ineffective in the upper layers, due to the relatively low density that reduces the effectiveness of the aforementioned three-body reaction.

The reservoir of CO$_2$ remains almost unaltered during the evolution, apart from the tiny variations in the upper layers of the high-CRIR cases (\case{Case2} and \case{Case4} in \fig{fig4}). Carbon dioxide chemistry is mainly controlled by the cosmic-rays driven destruction channel ${\rm CO_2 + CR \to CO + O}$, effectively balanced by ${\rm CO + OH \to H + CO_2}$, and by two minor formation reactions that involve oxygen, namely ${\rm HCO + O\to CO_2 + H}$ and ${\rm HCO + O_2 \to CO_2 + OH}$. Note that the oxygen produced by the CRIR dissociation of CO$_2$ determines the formation of OH, a key ingredient for the formation of water.

Carbon dioxide and water formations compete for OH, respectively via ${\rm CO + OH \to CO_2 + H}$ and ${\rm H_2 + OH \to H_2O + H}$. Both CO and OH are formed via the destruction of CO$_2$ and H$_2$O by CRIR, thus obtaining two groups of balanced reactions linked by OH. This explains the pressure-dependent behavior of CO and water (more pronounced in the high-CRIR cases), that closer to the surface present the same abundances, while in the upper layers CO becomes the dominating species. In particular, in the upper layers the higher CRIR and the lower densities enhance the formation of CO and the destruction of water. This effect is further enhanced by the lower temperatures of the high-pressure cases, that reduce the effectiveness of the formation of water (cfr.~the upper layers of \case{Case1} with \case{Case3} and \case{Case2} with \case{Case4}).

Molecular oxygen presents a clear vertical gradient, determined by ${\rm O_2 + O + M \to O_3 + M}$, a three-body reaction, and ${\rm O_3 + HCO \to CO_2 + O_2 + H}$, that both play a role in the lower layers of every model, with their efficiency controlled by the different temperature gradients, and by the availability of oxygen from the CO$_2$ CRIR dissociation.

Molecular hydrogen shows almost no vertical gradient, apart from the decrease in the lower layers of the atmosphere. The main reactions are ${\rm H_2O + CR \to H_2 + O}$ and ${\rm H_2 + OH \to H + H_2O}$ in the upper layers, and analogously ${\rm HCO + HCO \to 2\,CO + H_2}$ and ${\rm H_2 + O \to H + OH}$ closer to the surface, the former reaction employing HCO from the three-body reactions also responsible of water formation.

\subsection{Temporal evolution of superficial and upper atmospheric layers}
\begin{figure}
 \centerline{\includegraphics[width=0.5\textwidth]{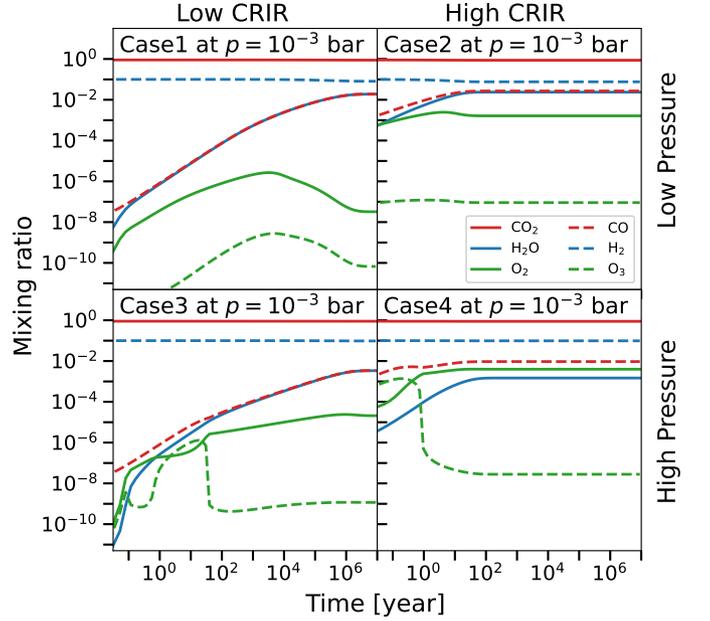}}
 \caption{Time evolution at $p=10^{-3}$~bar of the abundances of CO$_2$ (solid red), H$_2$O (solid blue), O$_2$ (solid green), CO (dashed red), and H$_2$ (dashed blue), and O$_3$ (dashed green), for the different models. Upper/lower row reports low/high-pressure cases, while the first/second column shows low/high-CRIR. This pressure value represents a typical layer from the outermost part of the atmosphere, see \fig{fig2}.}
 \label{fig5}
\end{figure}

\begin{figure}
 \centerline{\includegraphics[width=0.5\textwidth]{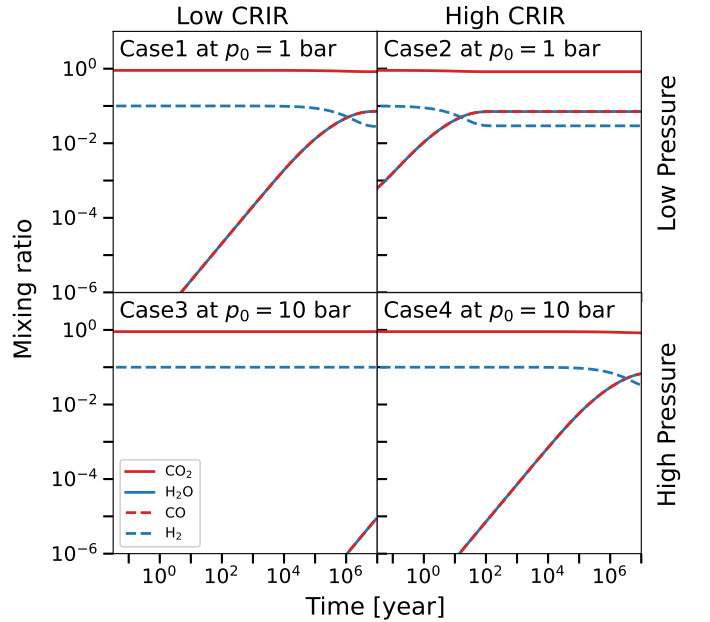}}
 \caption{Time evolution in the surface layer of the abundances of CO$_2$ (solid red), H$_2$O (solid blue), CO (dashed red), and H$_2$ (dashed blue), for the different models. Molecular oxygen and ozone are not reported having mixing ratios well below the scale of the plot. Upper/lower row reports low/high-pressure cases, while the first/second column shows low/high-CRIR.  For \case{Case1} and \case{Case2} the surface layer is $p=1$~bar, while for  \case{Case3} and \case{Case4} $p=10$~bar, see \fig{fig2}.}
 \label{fig6}
\end{figure}

In \fig{fig5} and \fig{fig6} we report the evolution in time for some of the chemical species, respectively for a layer at $p=10^{-3}$~bar and for the surface layer of each model ($p=1$ and $p=10$~bar). In the upper layers, CO$_2$ is unaffected during the evolution, as well as molecular hydrogen. The formation of water and CO is determined by the destruction of a part of CO$_2$, as discussed in the previous sections. CO is formed alongside water using part of CO$_2$ and H$_2$ that are both non-noticeably affected (note the log scale), apart from molecular hydrogen that slightly decreases with time in \fig{fig5} for \case{Case1} and \case{Case2}, as a consequence of the corresponding water formation. We note here that \case{Case3} reaches the chemical equilibrium around $10^6$~yr, considerably faster than the full system, which does not reach it even after $10^7$~yr (cfr.~\fig{fig3}, left panel). The time required to reach the chemical equilibrium is proportional to the CRIR that controls the kinetics, and hence the upper layers reach the equilibrium faster than the whole atmosphere that includes the surface layers, where CRIR are attenuated. Note that the equilibrium time-scale of the whole system reflects the time-scale of the lower layers, that comprise most of the total mass (see \fig{fig3}, right panel).

In the surface layer (\fig{fig6}) CO and water are always coupled, and a clear molecular-hydrogen-to-water conversion is present, as well as the self-similarity of the four cases. The CO-water coupling is determined by the same reasons discussed in the vertical profile Section, while the time scale of the molecular-hydrogen-to-water conversion is a direct consequence of the amount of CRIR and the different densities (note that \case{Case3} and \case{Case4} have a ten times higher pressure). In fact, \case{Case3}, where the CRIR have the largest attenuation, presents the slowest chemical kinetics, that does not reach the equilibrium even after $10^7$~yr. Conversely, in the high-CRIR and low-density \case{Case2}, the CRIR determines the shortest time-scale of the molecular-hydrogen-to-water conversion, since CRIR is less attenuated. Analogously, \case{Case1} and \case{Case4} have a similar behavior, since the CRIR has similar values closer to the surface (cfr.~\fig{fig2}).

These results show how the interplay between cosmic rays and the density structure of the moon determines the formation of water in the different layers and at different epoch (see~\fig{fig2}). In each case the total amount of water formed is similar, but with different time-scales (see~\fig{fig3}). While cosmic rays are crucial in determining the total abundance of species and to shape the chemistry of the middle and upper layers, in the ground layer(s) the density profile of the atmosphere plays a key role. In particular, an high-pressure and high-CRIR environment (\case{Case4}) mimics the behavior of a low-density and low-CRIR model (\case{Case1}), due to the similar cosmic rays attenuation (see e.g.~\fig{fig6}). These results suggest that the role of the variation in the temperature vertical profile is less relevant when compared to the cosmic rays attenuation effect, except in the upper layers, where the temperature affects the amount of water formed. However, note that contribution of the upper layers to the total mass is less relevant when compared to the lower layers (see \fig{fig3}).

\section{Discussion and Conclusion}\label{sect:discussion}
We modeled the time-dependent chemical evolution of the atmosphere of a 1~M$_\oplus$ exomoon orbiting around a 1~M$_{\rm J}$ free-floating planet. Rather than the radiation of the hosting star, CRIR is the main driver of the chemical kinetics, and the main source of heating are the tidal forces exerted by the planet onto its moon. We determined the amount of water produced in the CO$_2$-dominated atmosphere assuming an initial 10\% of H$_2$, and we measured the corresponding formation time-scale when changing the pressure at the base of the atmospheric vertical profile, the impinging CRIR, and the orbital parameters responsible for the tidal heating, namely semi-major axis and eccentricity.

Our findings suggest that a significant amount of water can be formed in the atmosphere of the exomoon and maintained in liquid form. In the low-pressure cases \case{Case1} and \case{Case2}, i.e.~the two models with the total atmospheric mass similar to Earth's, the amount of condensable water is of the order of $10^{17}$~kg at around 1~Myr and 100~yr, respectively. For comparison, the total mass of Earth's oceans is $\sim 1.4 \times 10^{21}$~kg \citep{weast1979}, and the amount of water vapor is $\sim10^{15}$~kg \citep{Trenberth2005}. The comet 67P/Churyumov-Gerasimenko has a total mass of $\sim10^{13}$~kg of which 10\%~water is a plausible estimate \citep{Choukroun2020}.

For the high-pressure model \case{Case4} we obtain a water mass of $1.27\times10^{18}$~kg at around 10~Myr. For comparison, on Venus, where the atmosphere is CO$_2$-dominated and approximately 10~times denser than our high-pressure cases, the mixing ratio of water at the surface is 20~ppm (smaller than our findings in general, excluding \case{Case3}), while the total mass of water in the atmosphere is $4\times10^{15}$~kg \citep{Svedhem2007}. However, on Venus the temperature and the pressure measured at the surface do not allow the presence of liquid water \citep{Way2020}. Mars also has a CO$_2$-dominated atmosphere, but considerably less massive than the atmosphere of Venus and hence not capable of any efficient greenhouse effect and to retain liquid water. The estimated amount of water in the Martian atmosphere is approximately $2\times10^{12}$~kg \citep{Jakosky1982}.

Given the boundary conditions, our results allow to classify this exomoon as a Class~II environment \citep{lammer-et-al2009}, corresponding to bodies that allow life, but differ from a Class~I habitat, i.e.~an Earth-like water-rich world. It is worth noticing that when comparing our model to Earth, we are ignoring any external mechanism of water accretion (e.g.~from asteroids, \citealt{Jin2019}), and we compare a CRIR-dominated with a radiation-dominated atmosphere, the latter deeply affecting the atmospheric chemistry (see e.g.~\citealt{hu-et-al2012}).

To maintain the temperature above the freezing point, the orbital parameters need to be constrained in time. As previously discussed, albeit surviving moons around ejected gas giants are expected to exist up to 0.1~au from the hosting planet, closer orbits ($\lesssim0.01$~au) are in general more probable \citep{rabago+steffen2019}. \fig{fig1} shows that, given the same pressure conditions, for the same eccentricity value, smaller semi-major axis implies warmer environments. However, the eccentricity is expected to decrease with time, toward a more circular orbit ($e\to0$), that corresponds to a reduction of the surface temperature, with a subsequent water solidification. This mechanism can be prevented by resonances \citep{debes+sigurdsson2007, hong-et-al2018} that are known to survive the ejection of the FFP from the hosting stellar system, allowing the stability of the orbital parameters over time, and hence the capability of the atmosphere to retain liquid water \citep{rabago+steffen2019}. A possible follow-up of our work consists in exploring the chemical evolution with the evolution of the orbital parameters.\\

The main conclusions of this work can be then summarized as follows:
\begin{itemize}[topsep=0pt]
    \item We found that an exomoon orbiting around a free-floating planet provides an environment that might sustain liquid water onto its surface if the optical thickness of the atmosphere is relatively large and the orbital parameters produce enough tidal heating to increase the temperature over the melting point of water. These orbital parameters are in agreement with previous simulations \citep{debes2007,debes+sigurdsson2007, hong-et-al2018, rabago+steffen2019}, and within the explored eccentricities (between $10^{-3}$ and 0.5) and semi-major axis (between $\sim10^{-3}$ and $\sim10^{-2}$~au).

    \item Due to the absence of impinging radiation, the timescale of water production is driven by the efficiency of cosmic rays in penetrating the atmosphere. Higher CRIRs reduce the water formation time-scale when compared to low-CRIR models, implying that they play a key role in the chemical evolution, by enhancing the chemical kinetics. However, due to the attenuation of cosmic rays, in the lower layers of the atmosphere, the water production is also affected by the density structure, that determines the integrated column density through the atmosphere. This causes an altitude-dependent abundance of water as well as of some of the other chemical species, as CO, H$_2$, and O$_2$.

    \item The temperature structure and its value at the surface are suitable for maintaining liquid water, but they play a marginal role in affecting the chemical evolution in the lower atmospheric layers (i.e.~the main contributors to the total water mass) when compared to the effects of CRIR and their attenuation caused by the vertical density profile.
\end{itemize}
\vspace{1mm}

\noindent The presence of water on the surface of the exomoon, affected by the capability of the atmosphere to keep a temperature above the melting point, might favor the development of prebiotic chemistry \citep{kasting93,stevenson1999}. According to \cite{stevenson1999}, a free-floating planet must have an effective temperature around 30~K to have liquid water on the surface. In our model the minimum effective temperature is $\sim$97~K for a surface pressure of 10~bar, in a CO$_2$-dominated atmosphere. The surface temperature is determined by the presence of greenhouse gases in the atmosphere (e.g.~CO$_2$) that control the heating produced by tidal forces. Under these conditions, if the orbital parameters are stable to guarantee a constant tidal heating, once water is formed, it remains liquid over the entire system evolution, and therefore providing favorable conditions for the emergence of life.

\section{Acknowledgements}
We thank the anonymous referees for improving the quality of this work.
SB acknolwedges support from the BASAL Centro de Astrofisica y Tecnologias Afines (CATA) AFB-17002.
This work was funded by the DFG Research Unit FOR 2634/1 ER685/11-1. SOD has been founded by Sophia University Special Grant for Academic Research and JSPS KAKENHI Grant Numbers 17H06105, 20H01975, and 26108511.
This research was supported by the Excellence Cluster ORIGINS, which is funded by the Deutsche Forschungsgemeinschaft (DFG, German Research Foundation) under Germany's Excellence Strategy - EXC-2094 - 390783311.
(\verb+http://www.universe-cluster.de/+).


\begin{thebibliography}{}

\bibitem[{{Badescu}(2010)}]{badescu2010}
\textbf{{Badescu} V} (2010) {Tables of Rosseland mean opacities for candidate
  atmospheres of life hosting free-floating planets}. \textit{Central European
  Journal of Physics} \textbf{8}, 463--479.
\bibitem[Badescu(2011{\natexlab{a}})]{badescu2011a}
\textbf{Badescu V} (2011) {Thermodynamic Constrains for Life Based on Non-Aqueous Polar Solvents on Free-Floating Planets}. \textit{Origins of Life and Evolution of the Biosphere} \textbf{41}, 73--99.

\bibitem[Badescu(2011{\natexlab{b}})]{badescu2011b}
\textbf{Badescu V} (2011) {Constraints on the free-floating planets supporting
 aqueous life}. \textit{Acta Astronautica} \textbf{69}, 788--808.

\bibitem[Badescu(2011{\natexlab{c}})]{badescu2011c}
\textbf{Badescu V} (2011) {Free-floating planets as potential seats for aqueous and non-aqueous life}.
\textit{Icarus} \textbf{216}, 485--491.

\bibitem[{{Barclay} \textit{et~al.}(2017)}]{barclay-et-al2017}
\textbf{{Barclay} T, {Quintana} EV, {Raymond} SN and {Penny} MT} (2017) {The
  Demographics of Rocky Free-floating Planets and their Detectability by
  WFIRST}. \textit{Astrophysical Journal} \textbf{841}, 86.

\bibitem[{{Barnes} and {O'Brien}(2002)}]{barnes+obrien2002}
\textbf{{Barnes} JW and O'Brien DP} (2002)
{Stability of Satellites around Close-in Extrasolar Giant Planets}.
\textit{The Astrophysical Journal} \textbf{575}, 1087--1093.

\bibitem[{Becker} \textit{et~al.}(2011)]{becker2011}
\textbf{{Becker} JK and {Black} JH and {Safarzadeh} M and {Schuppan} F} (2011) {Tracing the Sources of Cosmic Rays with Molecular Ions}. \textit{The Astrophysical Journal} \textbf{739:2}, L43.

\bibitem[{{Bennett} \textit{et~al.}(2014)}]{bennet-et-al2014}
\textbf{Bennett DP, Batista V, Bond IA, Bennet CS, Susuki D, Beaulieu JP, Udalski A, Donatowicz J, Bozza V, Abe F, Botzler CS,
Freeman M, Fukunaga D, Fukui A, Itow Y, Koshimoto N, Ling CH, Masuda K, Matsubara Y, Muraki Y, Namba S, Ohnishi K,
Rattenbury NJ, Saito To, Sullivan DJ, Sumi T, Sweatman WL, Tristram PJ, Tsurumi N, Wada K, Yock PCM, Albrow MD,
Bachelet E, Brillant S, Caldwell JAR, Cassan A, Cole AA, Corrales E, Coutures C, Dieters S, Dominis Prester D, Fouqué P,
Greenhill J, Horne J, Koo JR, Kubas D, Marquette JB, Martin R, Menzies JW, Sahu KC, Wambsganss J, Williams A, Zub M,
Choi JY, DePoy DL, Dong S, Gaudi BS, Gould A, Han C, Henderson CB, McGregor D, Lee CU, Pogge RW, Shin IG, Yee JC,
Szymański MK, Skowron J, Poleski R, Kozłowski S, Wyrzykowski Ł, Kubiak M, Pietrukowicz P, Pietrzyński G, Soszyński I,
Ulaczyk K, Tsapras Y, Street RA, Dominik M, Bramich DM, Browne P, Hundertmark M, Kains N, Snodgrass C, Steele IA,
Dekany I, Gonzalez OA, Heyrovský D, Kandori R, Kerins E, Lucas PW, Minniti D, Nagayama T, Rejkuba M, Robin AC and Saito R} (2014) {MOA-2011-BLG-262Lb: A Sub-Earth-Mass Moon Orbiting a
  Gas Giant Primary or a High Velocity Planetary System in the Galactic Bulge}.
  \textit{The astrophysical Journal} \textbf{785}, 155.


\bibitem[Caballero(2018)]{caballero2018}\textbf{Caballero JA} (2018)
{A review on substellar objects beyond the deuterium burning mass limit: planets, brown dwarfs or what?}.
\textit{Geosciences} \textbf{8}, 362--398.


\bibitem[{{Canup} and {Ward}(2006)}]{canup+ward2006}
\textbf{{Canup} RM and {Ward} WR} (2006) {A common mass scaling for satellite
systems of gaseous planets}. \textit{Nature} \textbf{441}, 834--839.


\bibitem[{{Catling} and {Zahnle}(2009)}]{catling+zahnle2009}
\textbf{{Catling} DC and {Zahnle} KJ} (2009) {The Planetary Air Leak}.
\textit{Scientific American} \textbf{300}, 36--43.


\bibitem[{{Choukroun}, {Altwegg} and {K\"uhrt}(2020)}]{Choukroun2020}
\textbf{{Choukroun} M, {Altwegg} K and {K\"uhrt} E} (2020) { Dust-to-Gas and Refractory-to-Ice Mass Ratios of Comet 67P/Churyumov-Gerasimenko from Rosetta Observations.}
\textit{Space Sci Rev} \textbf{216}, 44.


\bibitem[{{Clanton} and {Gaudi}(2016)}]{clanton+gaudi2016}
\textbf{{Clanton} C and {Gaudi} BS} (2016) {Synthesizing Exoplanet
  Demographics: A Single Population of Long-period Planetary Companions to M
  Dwarfs Consistent with Microlensing, Radial Velocity, and Direct Imaging
  Surveys}. \textit{Astrophysical Journal} \textbf{819}, 125.

\bibitem[{Clark} \textit{et al.}(2014)]{Clark2014}
\textbf{Clark RN, Swayze GA, Carlson R, Grundy W and Noll K} (2014) \textit{Spectroscopy from Space,Reviews in Mineralogy and Geo-chemistry}. Reviews in Mineralogy and Geo-chemistry, \textbf{78}, 399-–446

\bibitem[{{Dalgarno}(2006)}]{dalgarno2006}
\textbf{{Dalgarno} A} (2006) {Interstellar Chemistry Special Feature: The
  galactic cosmic ray ionization rate}. \textit{Proceedings of the National
  Academy of Science} \textbf{103}, 12269--12273.

\bibitem[{{Debes} and {Sigurdsson}(2007{\natexlab{a}})}]{debes2007}
\textbf{{Debes} JH and {Sigurdsson} S} (2007{\natexlab{a}}) {The Survival Rate
  of Ejected Terrestrial Planets with Moons}. \textit{Astrophysical Journal}
  \textbf{668}, 167--170.

\bibitem[{{Debes} and {Sigurdsson}(2007{\natexlab{b}})}]{debes+sigurdsson2007}
\textbf{{Debes} JH and {Sigurdsson} S} (2007{\natexlab{b}}) {The Survival Rate
  of Ejected Terrestrial Planets with Moons}. \textit{The Astropysical Journal Letters} \textbf{668},
  167--170.

\bibitem[Dobos \textit{et~al.}(2017)]{dobos-et-al2017}
\textbf{{Dobos} V, {Heller} R and {Turner} EL} (2017)
{The effect of multiple heat sources on exomoon habitable zones}.
\textit{Astronomy and Astrophysics} \textbf{601}, A91.

\bibitem[Dye (2012)]{dye2012}
\textbf{Dye ST} (2012) {Geoneutrinos and the radioactive power of the Earth}
\textit{Reviews of Geophysics} \textbf{50}, 3007-3026.

\bibitem[Fogg(2002)]{fogg2002}\textbf{Fogg MJ} (2002)
{Free-floating planets: their origin and distribution}.
\textit{MSc Astrophysics Project} \textbf{August 2002}, 1--54.

\bibitem[Fogg(1990)]{fogg1990}\textbf{Fogg MJ} (1990)
{Interstellar planets}. \textit{Communications in Astrophysics} \textbf{14}, 357--378.

\bibitem[Fox and Wiegert(2021)]{fox+wiegert2021}\textbf{{Fox} C and {Wiegert} P} (2021)
{MNRAS}. \textit{Exomoon candidates from transit timing variations: eight Kepler systemswith TTVs explainable by photometrically unseen exomoons} \textbf{510}, 2378--2393.

\bibitem[{Glein}(2015)]{Glein2015}
\textbf{{Glein} CR} (2015) \textit{Noble gases, nitrogen, and methane from the deepinterior to the atmosphere of Titan}. Icarus, \textbf{250}, 570--586


\bibitem[{{Guillot}(2010)}]{guillot2010}
\textbf{{Guillot} T} (2010) {On the radiative equilibrium of irradiated
  planetary atmospheres}. \textit{Astronomy and Astrophysics} \textbf{520},
  A27.


\bibitem[{{Hansen}(2008)}]{hansen2008}
\textbf{{Hansen} BMS} (2008) {On the Absorption and Redistribution of Energy in
Irradiated Planets}. \textit{The Astrophysical Journal Supplement Series}
\textbf{179}, 484--508.

\bibitem[Haqq-Misra and Heller (2018)]{haqq-misra2018}
\textbf{{Haqq-Misra} J and {Heller} R} (2018) {Exploring exomoon atmospheres with an idealized general circulation model}.
\textit{Monthly Notices of the RAS} \textbf{479}, 3477--3489.


\bibitem[Heller(2012)]{heller2012}
\textbf{Heller R} (2012) {Exomoon habitability constrained by energy flux and orbital stability}.
\textit{Astronomy and Astrophysics} \textbf{545}, L8.

\bibitem[{{Heller} and {Barnes}(2013)}]{heller+barnes2013}
\textbf{{Heller} R and {Barnes} R} (2013) {Exomoon Habitability Constrained by
Illumination and Tidal Heating}. \textit{Astrobiology} \textbf{13}, 18--53.

\bibitem[{{Heller} and {Zuluaga}(2013)}]{Heller2013}
\textbf{{Heller} R and {Barnes} R} (2013) {Magnetic shielding of exomoons beyond the circumplanetary habitable edge}. \textit{ApJL} \textbf{776}, L33.

\bibitem[Heller and Barnes(2014)]{heller+barnes2014}
\textbf{Heller R and Barnes R} (2014) {Constraints on the Habitability of Extrasolar Moons}.
\textit{Formation, Detection, and Characterization of Extrasolar Habitable Planets} \textbf{293}, 159--164.

\bibitem[{Heller} \textit{et~al.}(2014)]{heller-et-al2014}
\textbf{{Heller} R, {Williams} D, {Kipping} D, {Limbach} M, {Turner} E, {Greenberg} R,
{Sasaki} T, {Bolmont} E, {Grasset} O, {Lewis} K, {Barnes} R and {Zuluaga} J} (2014)
{Formation, Habitability, and Detection of Extrasolar Moons}.
\textit{Astrobiology} \textbf{14}, 1--42.

\bibitem[{Heller} \textit{et~al.}(2019)]{heller-et-al2019}
\textbf{{Heller} R, {Rodenbeck} K and {Bruno} G} (2019) {An alternative interpretation of the exomoon candidate signal in the combined Kepler and Hubble data of Kepler-1625}. \textit{Astronomy and Astrophysics} \textbf{624}, A95.

\bibitem[Henderson(2016)]{henderson2016}
\textbf{Henderson CB} (2016) {Using K2 to Find Free-floating Planets}.
\textit{American Astronomical Society Meeting Abstracts} \textbf{227}, 122.09.

\bibitem[{{Henning} \textit{et~al.}(2009)}]{henning-et-al2009}
\textbf{{Henning} WG, {O'Connell} RJ and {Sasselov} DD} (2009) {Tidally Heated
  Terrestrial Exoplanets: Viscoelastic Response Models}. \textit{Astrophysical
  Journal} \textbf{707}, 1000--1015.


\bibitem[{{Hindmarsh} \textit{et~al.}(2009)}]{Hindmarsh2005}
\textbf{{Hindmarsh} AC, {Brown} PN, Grant KE,  Lee SL, Serban R,  Shumaker DE, and Woodward CS} (2005) {{SUNDIALS}: Suite of nonlinear and differential/algebraic equation solvers}. \textit{ACM Transactions on Mathematical Software (TOMS)} \textbf{31}, 3

\bibitem[{{Hong} \textit{et~al.}(2018)}]{hong-et-al2018}
\textbf{{Hong} YC, {Raymond} SN, {Nicholson} PD and {Lunine} JI} (2018)
 {Innocent Bystanders: Orbital Dynamics of Exomoons During Planet-Planet
  Scattering}. \textit{Astrophysical Journal} \textbf{852}, 85.

\bibitem[Hu \textit{et~al}(2012)]{hu-et-al2012}
\textbf{Hu R, Seager S, Bains W} (2012) {Photochemistry in Terrestrial Exoplanet Atmospheres. I. Photochemistry Model and Benchmark Cases}. \textit{The Astrophysical Journal} \textbf{761}, 166.

\bibitem[{Jin and Bose (2019)}]{Jin2019}
\textbf{Jin Z and Bose M} (2019) \textit{New clues to ancient water on Itokawa}.
Science Advances \textbf{5}, 5.

\bibitem[{Jaupart \textit{et~al.}(2007)}]{jaupart-et-al2007}
\textbf{Jaupart C, Labrosse S and Mareschal JC} (2007) \textit{Treatise on geophysics}.
J. Geophys. Res., \textbf{87}, B4

\bibitem[{Jakosky \textit{et~al.}(1982)}]{Jakosky1982}
\textbf{Jakosky BM and Farmer CB} (1982) \textit{The seasonal and global behavior of water vapor in the Mars atmosphere: Complete global results of the Viking Atmospheric Water Detector Experiment}.
Elsevier Science, pp. 1--86.

\bibitem[{Kasting}(1982)]{Kasting1982}
\textbf{{Kasting} JF} (1982) \textit{Stability of ammonia in the primitive terrestrial atmosphere}. JGR Oceans, \textbf{87}, 3091--3098

\bibitem[{{Kasting} \textit{et~al.}(1993)}]{kasting93}
\textbf{{Kasting} JF, {Whitmire} DP and {Reynolds} RT} (1993) {Habitable Zones
  around Main Sequence Stars}. \textit{Icarus} \textbf{101}, 108--128.

\bibitem[Kreidberg \textit{et~al.}(2019)]{kreydberg2019}
\textbf{{Kreidberg} L, {Luger} R and {Bedell} M} (2019) {No Evidence for Lunar Transit in New Analysis of Hubble Space Telescope Observations of the Kepler-1625 System}. 
\textit{Astrophysical Journal Letters} \textbf{877}, L15.

\bibitem[Kipping \textit{et~al.}(2012)]{kipping-et-al2012}
\textbf{Kipping D, Bakos G, Buchhave L, Nesvorn\'y D and Schmitt A} (2012).
{The Hunt for Exomoons with Kepler (HEK): I. Description of a NewObservational Project}.
\textit{The Astrophysical Journal} \textbf{750}. 115-134.

\bibitem[{{Lammer} \textit{et~al.}(2009)}]{lammer-et-al2009}
\textbf{{Lammer} H, {Bredeh\"{o}ft} JH and {Coustenis} A} (2009) {What makes a planet habitable?}.
\textit{The Astronomy and Astrophysics Review} \textbf{17}, 181--249.

\bibitem[{{Lammer} \textit{et~al.}(2014)}]{lammer-et-al2014}
\textbf{{Lammer} H, {Schiefer} SC, {Juvan} I, {Odert} P, {Erkaev} NV, {Weber}
  C, {Kislyakova} KG, {G\"udel} M, {Kirchengast} G and {Hanslmeier} A} (2014)
  {Origin and Stability of Exomoon Atmospheres: Implications for Habitability}.
  \textit{Origins of Life and Evolution of the Biosphere} \textbf{44},
  239--260.

\bibitem[{{Lammer} \textit{et~al.}(2018)}]{lammer-et-al2018}
\textbf{{Lammer} H, {Zerkle} AL, {Gebauer} S, {Tosi} N, {Noack} L, {Scherf} M,
  {Pilat-Lohinger} E, {G{\"u}del} M, {Grenfell} JL, {Godolt} M and {Nikolaou}
  A} (2018) {Origin and evolution of the atmospheres of early Venus, Earth and
  Mars}. \textit{Astronomy and Astrophysics Reviews} \textbf{26}, 2.

\bibitem[{{Laughlin} and {Adams}(2000)}]{laughlin-adams2000}
\textbf{{Laughlin} G and {Adams} FC} (2000) {The Frozen Earth: Binary
  Scattering Events and the Fate of the Solar System}. \textit{Icarus}
  \textbf{145}, 614--627.

\bibitem[{{Lebrun} \textit{et~al.}(2013)}]{lebrun-et-al2013}
\textbf{{Lebrun} T, {Massol} H, {Chassefi{\`e}Re} E, {Davaille} A, {Marcq} E,
  {Sarda} P, {Leblanc} F and {Brandeis} G} (2013) {Thermal evolution of an
  early magma ocean in interaction with the atmosphere}. \textit{Journal of
  Geophysical Research (Planets)} \textbf{118}, 1155--1176.

\bibitem[{{Lehmer} \textit{et~al.}(2017)}]{lehmer-et-al2017}
\textbf{{Lehmer} OR, {Catling} DC and {Zahnle} KJ} (2017)
{The Longevity of Water Ice on Ganymedes and Europas around Migrated Giant Planets}.
\textit{The Astrophysical Journal} \textbf{839}, 32--41.

\bibitem[{{Lissauer}(1987)}]{lissauer1987}
\textbf{{Lissauer} JJ} (1987) {Timescales for planetary accretion and the
  structure of the protoplanetary disk}. \textit{Icarus} \textbf{69}, 249--265.

\bibitem[Liu \textit{et~al.}(2013)]{liu-et-al2013}
\textbf{{Liu} MC, {Magnier} EA,{Deacon} NR, {Allers} KN, {Dupuy} TJ, {Kotson} MC, {Aller} KM, {Burgett} WS,
{Chambers} KC, {Draper} PW, {Hodapp} KW, {Jedicke} R, {Kaiser} N, {Kudritzki} RP, {Metcalfe} N, {Morgan} JS,
{Price} PA, {Tonry} JL and {Wainscoat} RJ} (2013) {The Extremely Red, Young L Dwarf PSO J318.5338-22.8603:
A Free-floating Planetary-mass Analog to Directly Imaged Young Gas-giant Planets}.
\textit{The Astrophysical Journal Letters} \textbf{777}, 20--27.

\bibitem[Liu \textit{et~al.}(2016)]{liu2016}
\textbf{{Liu} MC, {Dupuy} TJ and {Allers} KN} (2016) {The Hawaii Infrared Parallax Program. II. Young Ultracool Field Dwarfs}.
\textit{The Astrophysical Journal} \textbf{833}, 96--161.

\bibitem[Luhman (2014)]{luhman2014}
\textbf{{Luhman} KL} (2014) {Discovery of a \raisebox{-0.5ex}\textasciitilde250 K Brown Dwarf at 2 pc from the Sun}.
\textit{The Astrophysical Journal Letters} \textbf{786}, L18.

\bibitem[{Mandt} \textit{et al.}(2014)]{Mandt2014}
\textbf{Mandt KE, Mousis O, Lunine J and Gautier D} (2014) \textit{Protosolar Ammonia as the Unique Source of Titan's nitrogen}. The Astrophysical Journal Letters , \textbf{788}, L24


\bibitem[{{Marley} and {Robinson}(2015)}]{marley+robinson2015}
\textbf{{Marley} MS and {Robinson} TD} (2015) {On the Cool Side: Modeling the
  Atmospheres of Brown Dwarfs and Giant Planets}. \textit{Annual Review of
  Astronomy and Astrophysics} \textbf{53}, 279--323.

\bibitem[{{Massol} \textit{et~al.}(2016)}]{massol-et-al2016}
\textbf{{Massol} H, {Hamano} K, {Tian} F, {Ikoma} M, {Abe} Y, {Chassefi{\`e}re}
  E, {Davaille} A, {Genda} H, {G{\"u}del} M, {Hori} Y, {Leblanc} F, {Marcq} E,
  {Sarda} P, {Shematovich} VI, {St{\"o}kl} A and {Lammer} H} (2016) {Formation
  and Evolution of Protoatmospheres}. \textit{Space Science Reviews}
  \textbf{205}, 153--211.

\bibitem[Molina-Cuberos \textit{et~al.}(2002)]{MolinaCuberos2002}
\textbf{{Molina-Cuberos} GJ, {Lichtenegger} H, {Schwingenschuh} K, {L\'opez-Moreno} JJ, and {Rodrigo} R} (2002) {Ion-neutral chemistry model of the lower ionosphere of Mars}. \textit{J. Geophys. Res.,} \textbf{107}, E5.

\bibitem[Mr\'oz \textit{et~al.}(2018)]{mroz-et-al2018}
\textbf{Mr\'oz P, Ryu YH, Skowron J, Udalski A, Gould A, Szymanski MK, Soszynsk I, Poleski R, Pietrukowicz P,
Kozlowski S, Pawlak M, Ulaczyk K, Albrow MD, Chung SJ, Jung YK, Han C, Hwang KH, Shin IG, Yee JC, Zhu W, Cha SM,
Kim DJ, Kim HW, Kim SL, Lee CU, Lee DJ, Lee Y , Park BG and Pogge RW} (2018) {A Neptune-mass Free-floating Planet Candidate Discovered by Microlensing Surveys}. \textit{The Astronomical Journal} \textbf{155}, 121--127.


\bibitem[Mr\'oz \textit{et~al.}(2020)]{mroz-et-al2020}
\textbf{Mr\'oz P, {Poleski} R, {Han} C, {Udalski} A, {Gould} A, {Szymanski} MK, {Soszynski} I, {Pietrukowicz} P, {Kozlowski} S, {Skowron} J, {Ulaczyk} K, {Gromadzki} M, {Rybicki} K, {Iwanek} P, {Wrona} M, {Albrow} MD, {Chung} S, {Hwang} K, {Ryu} Y, {Jung} YK, {Shin} I, {Shvartzvald} Y, {Yee} JC, {Zang} W, {Cha} S, {Kim} D, {Kim} H,
{Kim} S, {Lee} C, {Lee} D, {Lee} Y, {Park} B and {Pogge} RW} (2020) {A free-floating or wide-orbit planet in the microlensing event OGLE-2019-BLG-0551} \textit{arXiv e-prints}, arXiv:200301126.

\bibitem[{Murray and Dermott(2000)}]{murray+dermott2000}
\textbf{Murray CD and Dermott SF} (2000) \textit{Solar System Dynamics}. Cambridge
  University Press, pp. 130--186.

\bibitem[{Nimmo} and {Pappalardo}(2016)]{Nimmo2016}
\textbf{{Nimmo} F and {Pappalardo} RT} (2016) \textit{Ocean worlds in the outer solar system}. Journal of Geophysical Research: Planets, \textbf{121}, 1378--1399


\bibitem[Nordheim, \textit{et~al.}(2019)]{Nordheim2019}
\textbf{Nordheim TA,  Jasinski JM, and Hand KP} (2019) \textit{Galactic Cosmic-Ray Bombardment of Europa's Surface}. ApJL
  \textbf{881} L29


\bibitem[\"OPik(1964)]{opik1964}\textbf{\"Opik EJ} (1964)
{Stellar planets and little dark stars as possible seats of life}
\textit{Irish Astronomical Journal} \textbf{6}, 290--296.

\bibitem[{{Parmentier} and {Guillot}(2014)}]{parmentier+guillot2014}
\textbf{{Parmentier} V and {Guillot} T} (2014) {A non-grey analytical model for
  irradiated atmospheres. I. Derivation}. \textit{Astronomy and Astrophysics}
  \textbf{562}, A133.

\bibitem[{{Pollack} and {Yung}(1980)}]{pollack+yung1980}
\textbf{{Pollack} JB and {Yung} YL} (1980) {Origin and Evolution of Planetary
  Atmospheres}. \textit{Annual Review of Earth and Planetary Sciences}
  \textbf{8}, 425.

\bibitem[{{Rabago} and {Steffen}(2019)}]{rabago+steffen2019}
\textbf{{Rabago} I and {Steffen} JH} (2019) {Survivability of moon systems
  around ejected gas giants}. \textit{Monthly Notices of the RAS} \textbf{489},
  2323--2329.

\bibitem[Reynolds and Cassen(1978)]{reynolds1978}
\textbf{Reynolds RT and Cassen PM} (1978)
{Internal Structure of Large, Icy Satellites}.\textit{EOS Transactions} \textbf{59}, 1123.

\bibitem[{{Rimmer} and {Helling}(2013)}]{rimmer+helling2013}
\textbf{{Rimmer} PB and {Helling} C} (2013) {Ionization in Atmospheres of Brown
  Dwarfs and Extrasolar Planets. IV. The Effect of Cosmic Rays}.
  \textit{Astrophysical Journal} \textbf{774}, 108--124.

\bibitem[{{Rimmer} and {Helling}(2016)}]{rimmer+helling2016}
\textbf{{Rimmer} PB and {Helling} C} (2016) {A Chemical Kinetics Network for
  Lightning and Life in Planetary Atmospheres}. \textit{The Astrophysical
  Journal Supplement Series} \textbf{224}, 9--42.

\bibitem[{Robinson and Catling(2012)}]{robinson+catling2012}
\textbf{Robinson TD and Catling DC} (2012) An analytic radiative-convective
  model for planetary atmospheres. \textit{The Astrophysical Journal}
  \textbf{757}, 104--117.
  
\bibitem[Rodenbeck \textit{et~al.}(2018)]{rodenbeck2018}  
\textbf{{Rodenbeck} K, {Heller} R, {Hippke} M and {Gizon} L} (2018) {Revisiting the exomoon candidate signal around Kepler-1625 b}. \textit{Astronomy and Astrophysics} \textbf{617}, A49.
 
\bibitem[{Sagan(1969)}]{sagan1969}
\textbf{Sagan C} (1969) Gray and nongray planetary atmospheres structure,
  convective instability, and greenhouse effect. \textit{Icarus} \textbf{10},
  290--300.


\bibitem[Scharf(2006)]{scharf2006}
\textbf{Scharf CA} (2006) \textit{Instruments, Methods, and Missions for Astrobiology IX}.
San Diego: Society of Photo-optical Instrumentation Engineers, pp. 171--185.

\bibitem[Shapley(1958)]{shapley1958} \textbf{Shapley H} (1958)
\textit{Of stars and men. The human response to an expanding universe}.
London: Elek Books, p. 56.

\bibitem[Shapley(1962)]{shapley1962} \textbf{Shapley H} (1962)
{The scholar cornered: Crusted Stars and Self-Warming Planets}.
\textit{The American Scholar} \textbf{31}, 512--515.

\bibitem[{{Stevenson}(1999)}]{stevenson1999}
\textbf{{Stevenson} DJ} (1999) {Life-sustaining planets in interstellar space?}.
  \textit{Nature} \textbf{400}, 32.

\bibitem[{{Sumi} \textit{et~al.}(2011)}]{sumi-et-al2011}
\textbf{Sumi T, Kamiya K, Bennett DP, Bond IA, Abe F, Botzler CS, Fukui A, Furusawa K, Hearnshaw JB, Itow Y, Kilmartin PM,
Korpela A, Lin W, Ling CH, Masuda K, Matsubara Y, Miyake N, Motomura M, Muraki Y, Nagaya M, Nakamura S, Ohnishi K,
Okumura T, Perrot YC, Rattenbury N, Saito To, Sako T, Sullivan DJ, Sweatman WL, Tristram PJ, Yock PCM, Szymanski MK,
Kubiak M, Pietrzynski G, Poleski R, Soszynski I, Wyrzykowski L and Ulaczyk K} (2011)
  {Unbound or distant planetary mass population detected by gravitational
  microlensing}. \textit{Nature} \textbf{473}, 349--352.

\bibitem[{{Svedhem} \textit{et~al.}(2007)}]{Svedhem2007}
\textbf{Svedhem H, Titov D, Taylor F and Witasse O} (2007)
  {Venus as a more Earth-like planet}. \textit{Nature} \textbf{450}, 629--632.

\bibitem[Teachey \textit{et~al.}(2018)]{teachey-et-al2018}
\textbf{Teachey A, Kipping DM and Schmitt AR} (2018)
{HEK. VI. On the Dearth of Galilean Analogs in Kepler, and the Exomoon Candidate Kepler-1625b I}.
\textit{The Astronomical Journal} \textbf{155}, 36.

\bibitem[Teachey and Kipping(2018)]{teachey+kipping2018}
\textbf{Teachey A and Kipping DM} (2018)
{Evidence for a large exomoon orbiting Kepler-1625b}. \textit{Science Advances} \textbf{4}, eaav1784.


\bibitem[Trenberth and Smith (2005)]{Trenberth2005}
\textbf{Trenberth KE and Smith L} (2018)
{The Mass of the Atmosphere: A Constraint on Global Analyses}. \textit{Journal of Climate} \textbf{18(6)}, 864-875

\bibitem[{{Tsai} \textit{et~al.}(2017)}]{tsai-et-al2017}
\textbf{{Tsai} SM, {Lyons} JR, {Grosheintz} L, {Rimmer} PB, {Kitzmann} D and {Heng} K} (2017)
{VULCAN: An Open-source, Validated Chemical Kinetics Python
Code for Exoplanetary Atmospheres}. \textit{The Astrophysical Journal
Supplement Series} \textbf{228}, 20--46.

\bibitem[{Wakelam \textit{et~al.}(2015)}]{Wakelam2015}
\textbf{Wakelam V, Loison JC, Herbst E, Pavone B, Bergeat A, Béroff K, Chabot M, Faure A, Galli D, Geppert WD, Gerlich D, Gratier
P, Harada N, Hickson KM, Honvault P, Klippenstein SJ, Le Picard SD, Nyman G, Ruaud M, Schlemmer S, Sims IR, Talbi D,
Tennyson J and Wester R} (2015) \textit{The 2014 KIDA Network for Interstellar Chemistry}.
    The Astrophysical Journal Supplement Series, \textbf{217}, 20

\bibitem[{Wallace and Hobbs(2006)}]{wallace+hobbs2006}
\textbf{Wallace J and Hobbs P} (2006) \textit{Atmospheric Science: An Introductory Survey}.
Londres: Academic press, pp. 86.

\bibitem[Way and Del Genio(2020)]{Way2020} \textbf{Way, MJ, Del Genio, AD} (2020) \textit{Venusian Habitable Climate Scenarios: Modeling Venus Through Time and Applications to Slowly Rotating Venus-Like Exoplanets.} Journal of Geophysical Research (Planets) \textbf{125}(5), e2019JE006276

\bibitem[{{Weast} (1979)}]{weast1979} \textbf{{Weast} RC} (1979) \textit{CRC Handbook of chemistry and physics. A ready-reference book of chemical and physical data}.  Ohio: Chemical Rubber Comp. Press.

\bibitem[{{Weaver} and {Ramanathan}(1995)}]{weaber+ramanathan1995}
\textbf{{Weaver} CP and {Ramanathan} V} (1995) {Deductions from a simple
  climate model: Factors governing surface temperature and atmospheric thermal
  structure}. \textit{Journal of Geophysical Research} \textbf{100},
  11585--11592.

\bibitem[{Williams(2013)}]{williams2013}
\textbf{{Williams} DM} (2013) {Capture of Terrestrial-Sized Moons by
 Gas Giant Planets}. \textit{Astrobiology}
\textbf{13}, 315--323.


\bibitem[{{Yoder} and {Peale}(1981)}]{yoder+peale1981}
\textbf{{Yoder} CF and {Peale} SJ} (1981) {The tides of Io}. \textit{Icarus}
\textbf{47}, 1--35.

\bibitem[Zapatero Osorio \textit{et~al.}(2000)]{zapatero-et-al2000}
\textbf{Zapatero Osorio MR, {B{\'e}jar} VJS, {Mart{\'\i}n} EL, {Rebolo} R,
{Barrado y Navascu{\'e}s} D, {Bailer-Jones} CAL and {Mundt} R} (2000) {Discovery of Young Isolated Planetary Mass Objects in the {\ensuremath{\sigma}} Orionis Star Cluster}. \textit{Science} \textbf{290}, 103--107.


\bibitem[Zhong and Haigh(2013)]{Zhong2013}
\textbf{{Zhong} W and {Haigh}, JD} (2013) {The greenhouse effect and carbon dioxide}. \textit{Weather} \textbf{68}, 100--105.




\end{thebibliography}




\end{document}